\documentclass[
    ,final            % use final for the camera ready runs
  ]
  {aipproc}

\def\ergss{erg s$^{-1}$}
\def\Chandra{${\it Chandra}$}

\def\HST{${\it HST}$}
\newcommand{\Msun}{\ifmmode {M_{\odot}}\else${M_{\odot}}$\fi}
\newcommand{\Lsun}{\ifmmode {L_{\odot}}\else${L_{\odot}}$\fi}
\layoutstyle{8x11single}

\begin{document}

\title{X-ray Sources in Galactic Globular Clusters}

\classification{95.85.Nv, 97.10.gz, 97.60.Jd, 97.60.Gb, 97.80.Gm, 97.80.Jp, 98.20.Gm, 98.70.Qy}
%\classification{<Replace this text with PACS numbers; choose from this list:
%                \texttt{http://www.aip.org/pacs/index.html}>}
\keywords      {globular clusters, X-ray binaries, cataclysmic variables, pulsars}

\author{Craig O. Heinke}{
  address={Dept. of Physics, University of Alberta, Room 238 CEB, Edmonton, AB, T6G 2G7, Canada; heinke@ualberta.ca}
  ,altaddress={Ingenuity New Faculty} % additional visiting address
}

\begin{abstract}
 I review recent work on X-ray sources in Galactic globular clusters, identified with low-mass X-ray binaries (LMXBs), cataclysmic variables (CVs), millisecond pulsars (MSPs) and coronally active binaries by Chandra.  Faint transient LMXBs have been identified in several clusters, challenging our understanding of accretion disk instabilities.  Spectral fitting of X-rays from quiescent LMXBs offers the potential to constrain the interior structure of neutron stars.  The numbers of quiescent LMXBs scale with the dynamical interaction rates of their host clusters, indicating their dynamical formation.  Large numbers of CVs have been discovered, including a very faint population in NGC 6397 that may be at or beyond the CV period minimum.  Most CVs in dense clusters seem to be formed in dynamical interactions, but there is evidence that some are primordial binaries.  Radio millisecond pulsars show thermal X-rays from their polar caps, and often nonthermal X-rays, either from magnetospheric emission, or from a shock between the pulsar wind and material still flowing from the companion.  Chromospherically active binaries comprise the largest number of X-ray sources in globular clusters, and their numbers generally scale with cluster mass, but their numbers seem to be reduced in all globular clusters compared to other old stellar populations.
\end{abstract}

\maketitle

%%%%%%%%%%%%%%%%%%%%%%%%%%%%%%%%%%%%%%%%%%%%
%% MAINMATTER
%%%%%%%%%%%%%%%%%%%%%%%%%%%%%%%%%%%%%%%%%%%%

This article briefly reviews key X-ray observations and related work on Galactic globular clusters, concentrating on work published since 2005.  A more complete, but now dated, review is \citet{Verbunt04}, which nicely covers earlier work.  See also \citet{Rasio07}, \citet{Verbunt08}, and \citet{Pooley10}.

\section{Low-Mass X-ray Binaries}

Observations by the first X-ray satellite Uhuru demonstrated a strong overabundance of X-ray sources in globular clusters compared to the rest of the galaxy, which was quickly attributed to dynamical formation mechanisms in the dense cores of globular clusters \citep{Clark75}.  The bright ($L_X>10^{36}$ \ergss) X-ray sources in globular clusters have mostly been shown to be neutron stars (NSs) in low-mass X-ray binaries (LMXBs) through their X-ray bursts \citep{Lewin95,intZand03}. 
 Several mechanisms to create these systems have been suggested, including exchange of NSs into primordial binaries \citep{Hills76}, tidal capture of another star by the NS \citep{Fabian75}, and collisions of NSs with giants \citep{Sutantyo75}.  These mechanisms all have a similar dependence on the structural parameters of the host globular clusters, as they depend on the rate of close encounters, and most encounters take place in the nearly-constant-density core of the clusters.  Comparisons of the close encounter rates between clusters  often use the parametrization $\Gamma=\rho_c^{2}r_c^{3}/\sigma=\rho_c^{1.5}r_c^2$ \citep{Verbunt87}, where $\rho_c$ is the central density, $r_c$ is the cluster core radius, and $\sigma$ is the central velocity dispersion (a lower $\sigma$ promotes gravitational focusing in close encounters).  Indeed, this quantity reasonably describes the probability of finding LMXBs in Galactic globular clusters \citep{Verbunt87}, and (although limited by our optical spatial resolution) is also related to the probability of finding LMXBs in globular clusters in other galaxies \citep{Jordan04,Sivakoff07, Jordan07,Peacock09}.

Fifteen bright LMXBs are now known in 12 globular clusters, of which at least seven are transients showing bright ``outbursts'' and longer periods in which they are quite faint \citep{Verbunt04,Heinke10}.  Periods are now known for ten of these systems, with recent discoveries of the periods of M15 X-2 \citep[22.6 min.,][]{Dieball05}, SAX J1748.9-2021 in NGC 6440 \citep[8.5 hrs,][]{Altamirano08}, 4U 0513-40 in NGC 1851 \citep[17 min.,][]{Zurek09}, NGC 6440 X-2 \citep[57.3 min.,][]{Altamirano10}, and IGR J17480-2446 in Terzan 5\citep[21.252 hours,][]{Papitto10,Strohmayer10}.  \citet{Deutsch00} pointed out the excess of ``ultracompact'' periods (now at least 5 of 15) below one hour among globular cluster sources (compared to the period distribution among LMXBs in the rest of the Galaxy), indicating degenerate companions.  Such systems are most easily created through collisions of NSs with red giants \citep[e.g.][]{Ivanova05}, indicating the importance of this mechanism.  

Two transient systems (both, coincidentally, in NGC 6440; Fig. 1) have been found to show coherent millisecond X-ray pulsations during outbursts, identifying the rotational period of the NSs \citep{Altamirano08,Altamirano10}, while the new transient in Terzan 5, IGR J17480-2446 \citep{Bordas10,Pooley10b}, is a slow (11 Hz) pulsar \citep{Strohmayer10b}.  NGC 6440 X-2 (and, so far, IGR J17480-2446) show pulsations throughout all outbursts, while SAX J1748.9-2021 shows them only occasionally, perhaps due to its higher average mass-transfer rate burying its magnetic field \citep{Cumming01}.  NGC 6440 X-2 shows unusual outburst behavior, with its outbursts lasting $<$4 days (above $10^{35}$ erg/s), reaching relatively low peak $L_X$s of $<2\times10^{36}$ erg/s, and (most unusually) recurring every $\sim$31 days during much of 2009 and 2010 \citep{Heinke10}.  As it is hard to identify such faint outbursts even in the sensitive RXTE PCA Galactic Bulge scans \citep{Swank01}, it seems quite likely that other such weak transients may be missed by current instruments.  

\begin{figure}
  \includegraphics[height=.36\textheight]{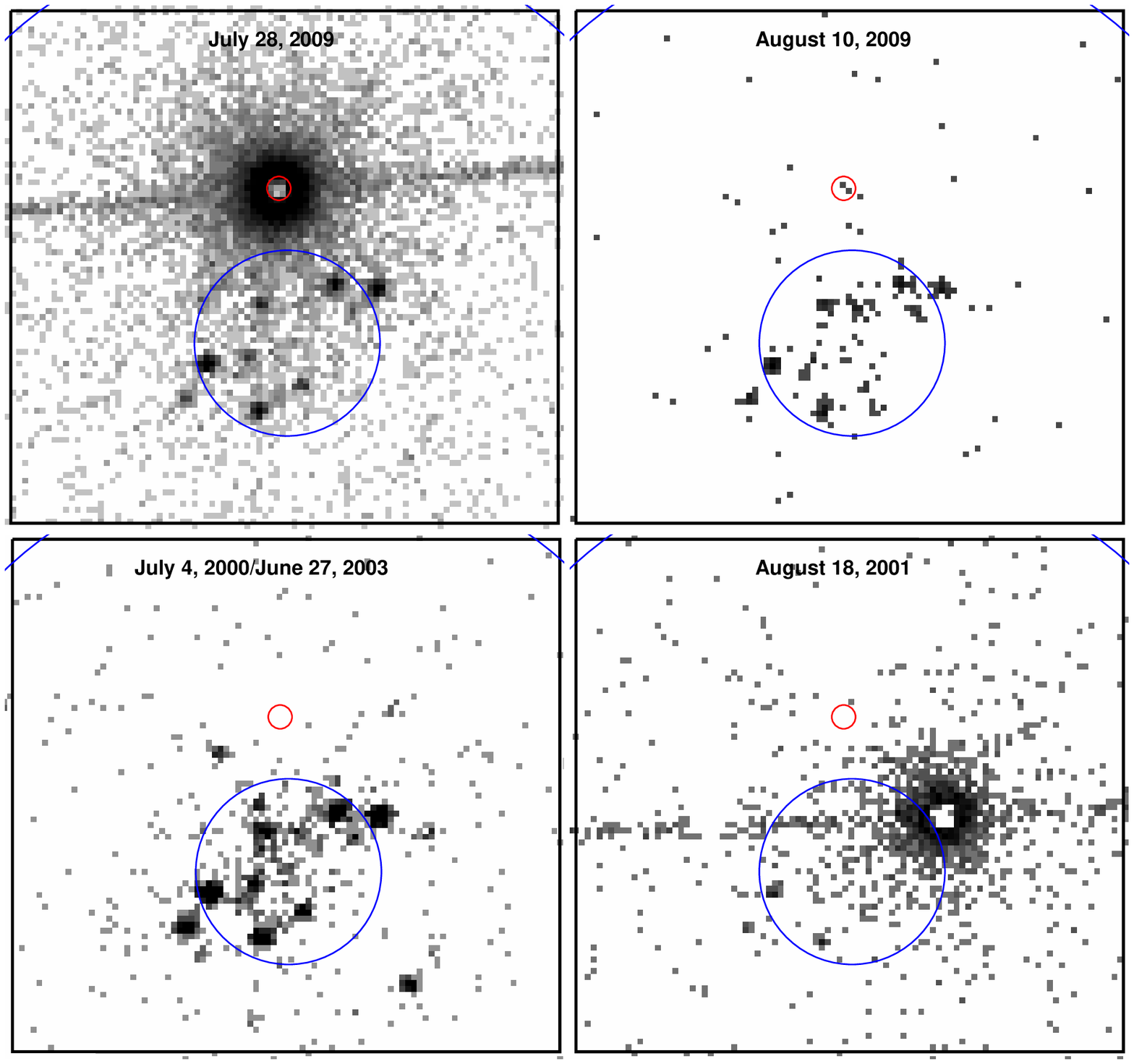}
  \includegraphics[height=.36\textheight]{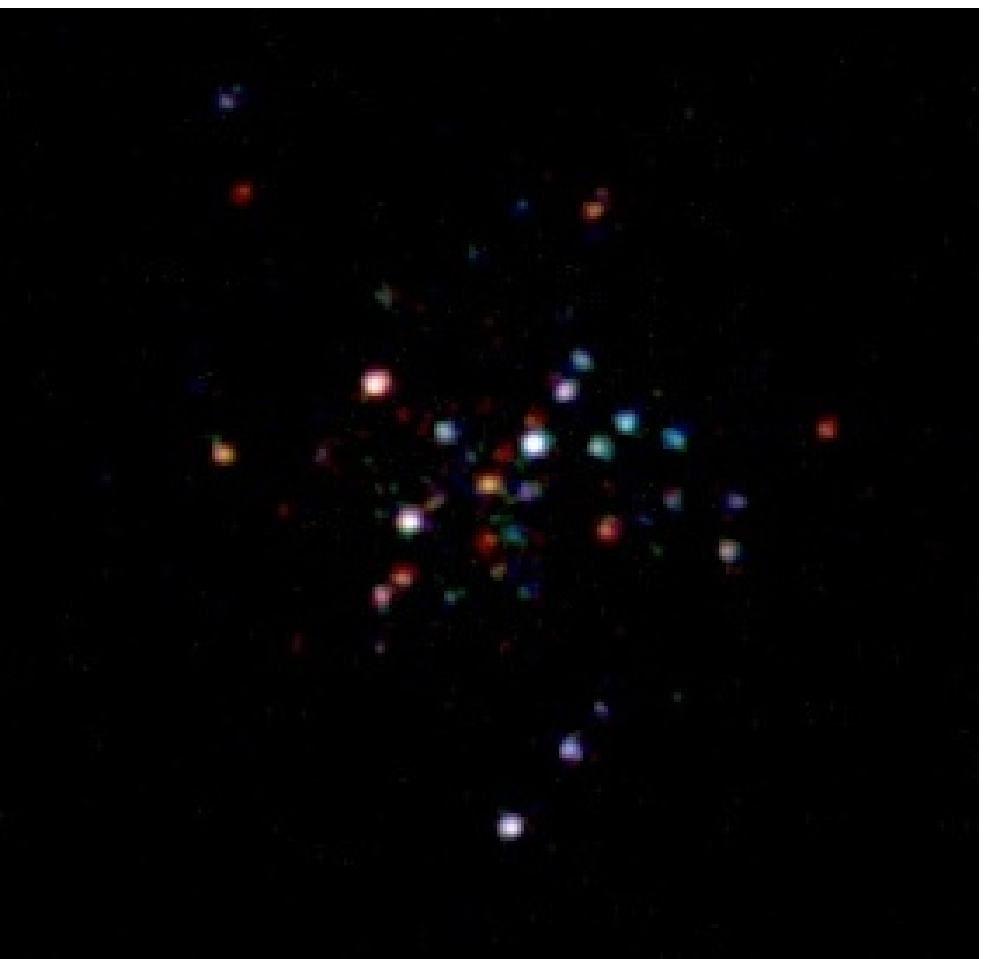}
  \caption{{\bf Left:} \Chandra\ X-ray images of the cluster NGC 6440 during an outburst of NGC 6440 X-2 (top left), one week later (top right), during an outburst of SAX J1748.9-2021 (bottom right), and from two quiescent periods (bottom left).  Images on the right have $\sim$1/10 the exposure time as those on the left.  The cluster core radius, and the position of 6440 X-2, are indicated on all panels \citep{Heinke10}.
{\bf Right:} Representative-color X-ray image of the extremely rich globular cluster Terzan 5 (1-2 keV red, 2-3 keV green, 3-6 keV blue).  Quiescent LMXBs generally appear reddish, while the white sources are likely dominated by cataclysmic variables.}
\end{figure}

Even fainter transients, the very faint X-ray transients ($10^{34}<L_{X,peak}<10^{36}$ \ergss), have been identified near the Galactic Center \citep[e.g.][]{Wijnands06}.  Some of the objects displaying such weak outbursts also show ``normal'' brighter  outbursts, while others have not shown ``normal'' outbursts, leading to speculation that they require unusual evolutionary histories with extremely low-mass companions \citep{King06}.  Searches for such behavior in globular clusters offer the opportunity to study these  objects in the optical/UV and soft X-ray, which is not feasible in the Galactic Center.  One X-ray source in the globular cluster M15 (M15 X-3) appeared at $L_X$(0.5-10 keV)$\sim6\times10^{33}$ \ergss in 1994-1995, and in 2004 and 2007, and at $L_X<10^{32}$ \ergss in 2000 and 2001 \citep{Heinke09b}.  A likely \HST\ counterpart (with a strong $U$ excess in one epoch) suggests a mass of $\sim$0.65 \Msun for the companion, ruling out some explanations for very faint transient behavior.    

\section{Quiescent LMXBs}

Transient LMXBs in quiescence typically are $\sim10^4$ times fainter than in outburst, showing X-ray spectra composed of a blackbody-like component and/or a spectrally harder component often fit with a power-law of photon index 1--2 \citep{Campana98a}.  The blackbody component is typically believed to be heat deposited in the neutron star deep crust during outbursts \citep{Brown98}, emerging through a hydrogen atmosphere on the surface \citep{Zavlin96}.   

Many quiescent LMXBs show spectra dominated by the blackbody-like component, allowing X-ray spectral identification of these systems in clusters \citep{Grindlay01a,Rutledge02a}.  An example of an X-ray color image of one such cluster, the rich globular Terzan 5 \citep{Heinke06b}, is shown in Fig. 1 (right, color online), where the quiescent LMXBs can often be distinguished from other sources by their X-ray colors.  

The numbers of quiescent LMXBs observed in different globular clusters are in accord with the predictions of the $\Gamma$ formalism \citep{Heinke03d,Heinke06b}, and with more sophisticated integrations of the collision probability per unit volume, $\rho^2/\sigma$, from the cluster centers to their half-mass radii \citep{Pooley03,Pooley06}.  Extrapolating to the full globular cluster system, roughly 200 quiescent LMXBs are predicted \citep{Heinke05b}. This indicates an average duty cycle of $<$3\% for transient LMXBs in globular clusters, and suggests an average recurrence time of $\sim$1000 years (if 7 of $\sim$200 transients have entered outburst over the $\sim$40 years of X-ray satellites).  However, if smaller outbursts (such as those from NGC 6440 X-2 or M15 X-3) are common, many neutron stars in quiescent LMXBs can be heated without ever producing major outbursts.

\begin{figure}
\includegraphics[height=.37\textheight]{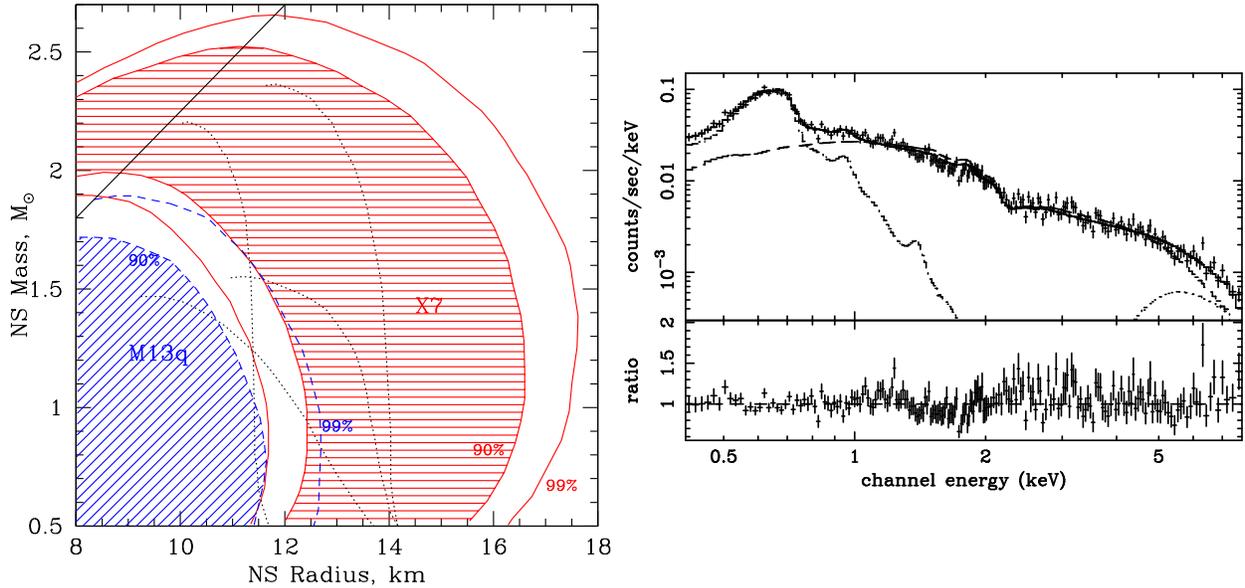} 
  \caption{{\bf Left:} Constraints on NS radius and mass from \Chandra\ observations of the NS X7 in 47 Tuc \citep{Heinke06a} and on the NS in M13 \citep{Webb07}.  The 90\% confidence regions for each are shaded, while the 99\% confidence regions are indicated by outer contour lines (solid for X7, dashed for the M13 NS). Some illustrative NS equations of state \citep{Lattimer04} are shown as dotted lines. 
{\bf Right:} Energy spectrum of the CV X9 in 47 Tuc, modeled with three thermal plasma components at 0.25, $>17$, and $>$6 keV, with the third component absorbed behind $N_H$=$9\times10^{23}$ cm$^{-2}$ \citep{Heinke05a}.}
\end{figure}

The well-known distances of globular clusters allow us to constrain a range in mass and radius (essentially, to constrain the radiation radii) of the NSs in quiescent LMXBs, by measuring the temperature and luminosity of the thermal radiation with \Chandra\ or XMM \citep{Rutledge02a}.  Only a few NSs, in 47 Tuc \citep{Heinke06a}, NGC 6397 \citep{Guillot10}, $\omega$ Cen and M13 \citep{Webb07} have constraints that exclude some considered NS equations of state \citep{Lattimer04}.  Intriguingly, the constraints on X7 in 47 Tuc and on the NS in M13 disagree at the 90\% confidence level, though not at the 99\% level (Fig. 2). Further studies of these and other NSs would be of great interest.

\section{Cataclysmic Variables}

Faint X-ray sources in globular clusters were discovered with {\it Einstein} and suggested to be mostly cataclysmic variables (CVs), white dwarfs accreting from low-mass companions \citep{Hertz83}.  Substantial populations have been predicted in globular clusters, but searches for CV outburst behavior have found very few outbursts \citep[e.g.][]{Shara96}.
A few X-ray sources were identified with blue, variable stars using ROSAT and \HST\ \citep{Cool95,Verbunt98}.  The accuracy of \Chandra\ positions, along with deep \HST\ optical/near-UV imaging, has allowed the identification of dozens of 
strong candidate CVs \citep{Grindlay01b,Pooley02a,Edmonds03a}. 

Since background AGN are blue and can produce high X-ray/optical flux ratios, they can masquerade as CVs \citep{Bassa05}, and an additional characteristic is needed to confidently identify CVs.  
Recent confident identifications of CVs also use H$\alpha$-$R$ color-magnitude diagrams (CMDs) to identify H-$\alpha$ emission \citep{Bassa08,Lu09}, colors closer to the main-sequence in optical CMDs than in near-UV CMDs \citep[indicative of a blue disk plus cool star;][]{Lugger07,Huang10}, or strong optical/UV variability \citep{Dieball07}.  An alternative approach to identifying X-ray sources uses imaging of clusters in the less crowded far-UV, where UV-luminous CVs have colors between those of white dwarfs and the main sequence \citep{Knigge02}.  X-ray detections \citep{Servillat08b,Dieball10} or identification of strong variability \citep{Dieball07} are usually necessary to ensure these objects are indeed CVs, as spectroscopy indicates that not all such ``gap'' sources are CVs \citep{Knigge08}.

Some CVs found in globular clusters show unusual X-ray behavior.  \citet{Bassa08} identify an extended hard X-ray source around a likely CV in NGC 6366, which they suggest may be an old nova remnant (no other nova remnants show such hard X-ray spectra).  \citet{Heinke05a} identify a strong low-energy (0.65 keV) spectral line in the CV X9 in 47 Tuc (Fig. 2a), which is also not understood. 

The CVs found in globular clusters seem to show fewer outbursts than known CVs in our local region of the Galaxy \citep[e.g.][]{Pietrukowicz08}, and higher X-ray/optical flux ratios \citep{Edmonds03b}.  The lack of outbursts has been explained as due to an preponderance of magnetic CVs (with truncated, thus more stable, accretion disks) \citep{Grindlay95}, low mass transfer rates \citep{Edmonds03b}, or a combination of both \citep{Dobrotka06}.  The high X-ray/optical flux ratios favor low mass transfer rates \citep{Edmonds03b}, but the high X-ray luminosities are then difficult to understand \citep{Verbunt97}.  A possible explanation is that higher-mass white dwarfs are preferentially exchanged into binaries in globular clusters \citep{Ivanova06}.  As more massive white dwarfs are denser, and tend to have higher B fields, this would tend toward higher X-ray/optical flux ratios and toward more magnetic CVs.
Recently, a large population of extremely faint ($M_V>10$) CVs has been uncovered through deep X-ray and optical imaging of NGC 6397 \citep[Fig. 3, left; ][]{Cohn10}, suggesting CVs near or beyond the period minimum, as recently identified in the field \citep[e.g.][]{Gansicke09,Patterson09}.

\begin{figure}
  \includegraphics[height=.47\textheight]{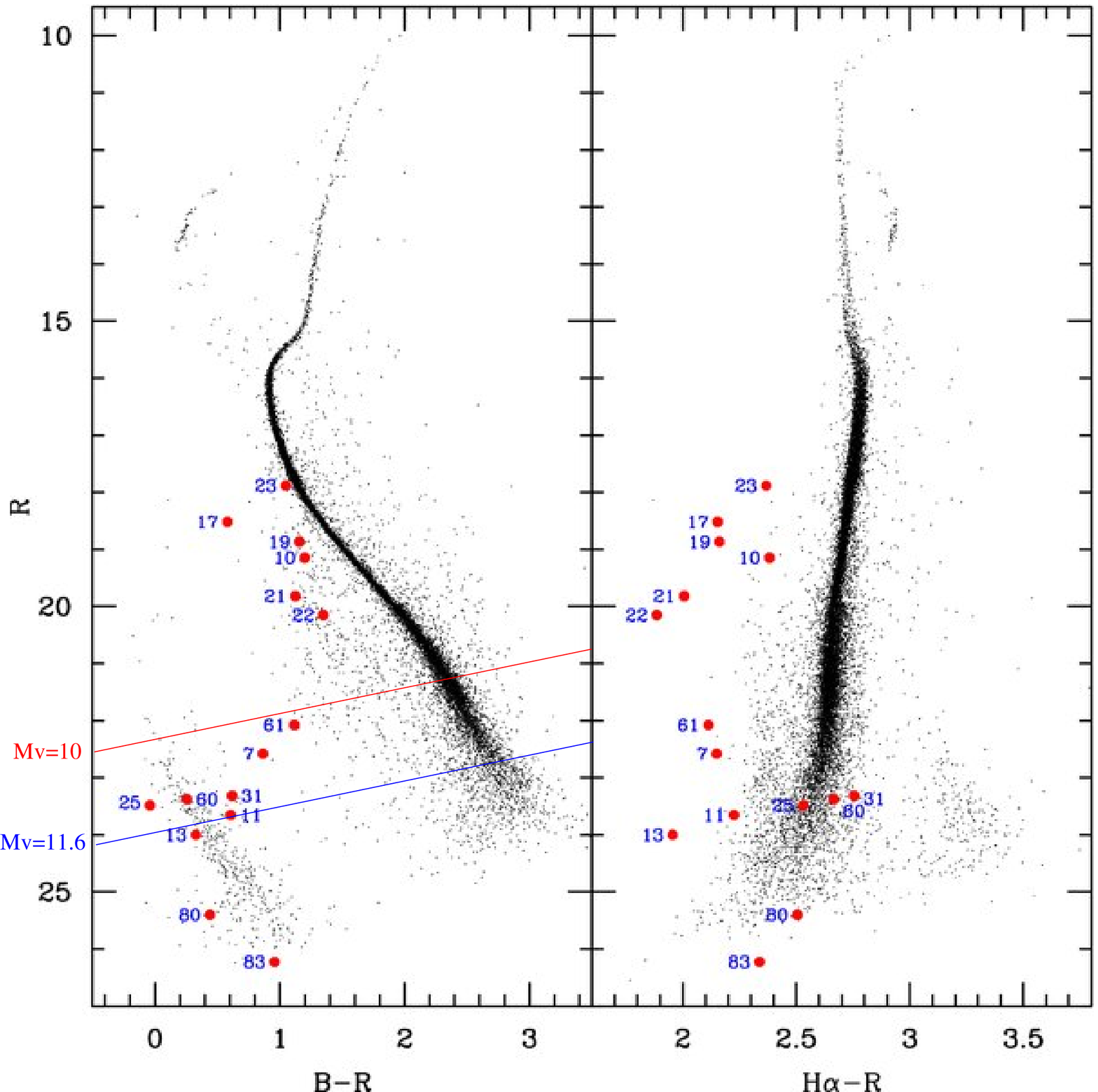}
  \includegraphics[height=.45\textheight]{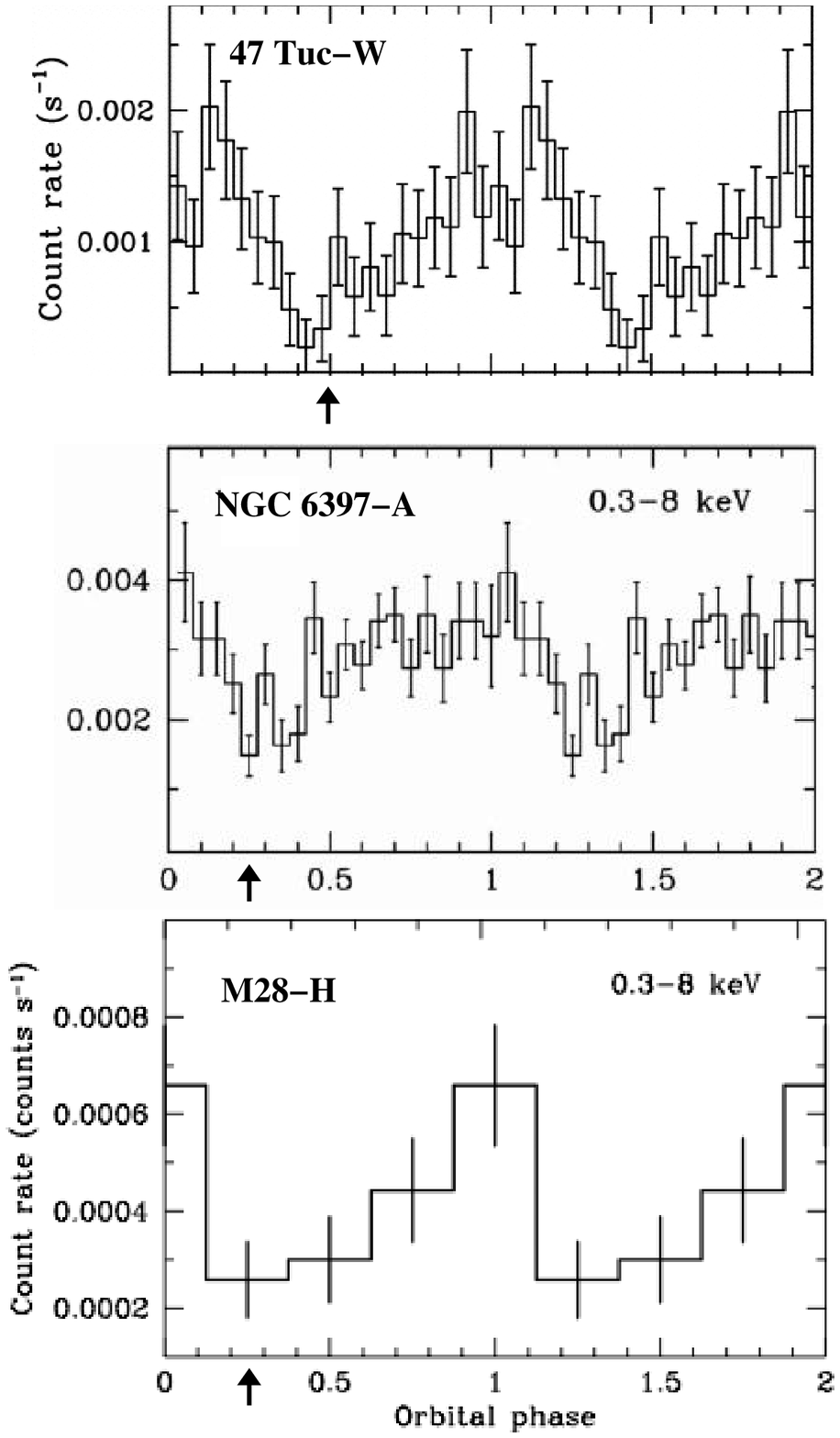}
  \caption{{\bf Left:} Color-magnitude diagrams of NGC 6397, with X-ray identified CVs indicated \citep{Cohn10}.  A new population of very faint, low-mass-transfer CVs is identified.  {\bf Right:} Phase-binned lightcurves of three low-mass eclipsing MSPs, with the times of superior conjunction of the NS indicated \citep{Bogdanov05,Bogdanov10a,Bogdanov10b}.  Each shows evidence for an X-ray eclipse, likely produced by the companion eclipsing an X-ray emitting shock between the pulsar wind and continuing mass transfer from the companion \citep{Bogdanov05}.} 
\end{figure}

A major current question about CVs is what fraction of them are produced dynamically, vs. being descended from primordial binaries through standard binary evolution.  By studying the distribution of CVs among globular clusters with different structures, this can be constrained, as for quiescent LMXBs (see above).  Despite substantial efforts (above), few clusters have sufficient optical data to identify the majority of X-ray sources, requiring alternate approaches.  Although not all cluster members with hard X-ray spectra and $L_X$ from   $10^{31}$--$10^{33}$ \ergss are CVs, the vast majority of these seem to be CVs.  \citet{Pooley06} and \citet{Heinke06b} showed that these hard X-ray sources are largely dynamically produced, though some fraction appear to be primordial in origin.  This is consistent with recent theoretical modeling of CVs in clusters \citep{Ivanova06,Shara06}.  These models also suggest that many binaries, that would otherwise become CVs, are destroyed in the high density environs of globular clusters.  Observations now confirm that the number,  and the total X-ray luminosity, of CVs per unit mass in low-to-moderate density globular clusters (such as $\omega$ Cen) is significantly reduced compared to the Galaxy as a whole and to lower-density open clusters \citep{Haggard09,Heinke10c}.  

\section{Millisecond Radio Pulsars}

Millisecond radio pulsars (MSPs) are the evolutionary products of low-mass X-ray binaries, and thus are also overabundant in globular clusters \citep[e.g.][]{Camilo05}, with $\sim$140 now known \citep{Ransom08}.  
A number of MSPs have now been identified in the X-ray, including all MSPs with known positions in 47 Tuc \citep{Heinke05a,Bogdanov06}.  Spectral studies of those MSPs indicate that the majority are well-described by thermal emission from the polar caps of NSs with hydrogen atmospheres \citep{Bogdanov06}.  A minority show nonthermal emission, which may be due to magnetospheric emission (as seen in pulsar M28A, \citet{Rutledge04, Bogdanov10b}), confusion with other X-ray sources, or shocks between the pulsar wind and other material \citep[][see below]{Bogdanov05}.  There is no evidence for substantial differences in the X-ray properties of MSPs in globular clusters vs. those in the rest of the Galaxy \citep{Bogdanov06}.

There is no evidence for a correlation between the X-ray luminosity and radio luminosity of the known MSPs in 47 Tuc \citep{Heinke05a}.  This suggests that any MSPs not yet detected in the radio should show X-ray characteristics like the known MSPs, and comparison of the known MSPs with unidentified X-ray sources permits a limit ($<60$, probably $\sim$25) on the total number of MSPs in 47 Tuc \citep{Heinke05a}.  A similar approach constrained the number of MSPs in NGC 6397 to $<$6 \citep{Bogdanov10a}.  
Assuming that the distribution of MSPs among globular clusters can be predicted by encounter rate (like their progenitors the LMXBs), and using the well-constrained number in 47 Tuc as a point of reference, of order 700 MSPs are predicted to exist in the Galactic globular cluster system  \citep{Heinke05a}.  

Some  MSPs with companions of $\sim$0.1-0.2 \Msun\ show radio eclipses, indicating the presence of gas escaping from a main-sequence companion star \citep{Freire04}.  Three of these systems--47 Tuc-W, NGC 6397-A, and M28-H, have now been carefully studied in the X-rays, and all show evidence for hard X-ray spectra and X-ray eclipses \citep[][see Fig. 3, right]{Bogdanov05,Bogdanov10a,Bogdanov10b}. 
 The X-ray eclipses are consistent in phase with the superior conjunction of the NS, but are far too wide in phase to be explained as an eclipse of the NS by the companion.  This suggests that the emission region is a shock between the pulsar wind and material flowing from a (Roche-lobe filling) companion star, located close to the companion star (allowing long eclipses) \citep{Bogdanov05}.  These objects illustrate a phase in the transition between LMXBs and MSPs, where the pulsar has turned on but material has not yet stopped flowing from the companion star. 

\section{Chromospherically Active Binaries}

Normal stars on the lower main sequence produce X-rays in their coronae, generated by a dynamo mechanism within their convective zone.  Their X-ray luminosity is determined by their rotation rate, so as they age their X-ray production drops to nondetectability at kpc distances.  However, stars in close binaries (``active binaries'', or ABs) are tidally locked, and thus forced to rotate at high rates, leading to strong X-ray production.  ABs make up the vast majority of the faintest ($L_X<10^{31}$) globular cluster X-ray sources.

ABs are typically identified in globular clusters by positional coincidence of a main-sequence star (or slightly above the main sequence) with an X-ray source.   There are two difficulties in identifying ABs; the star in the error circle could just be a chance coincidence (with the true counterpart being undetected), or the star might be a CV, active galactic nucleus (AGN), or foreground star.  A location on the cluster main sequence essentially rules out an AGN or foreground star nature (foreground stars are typically redder than the main sequence, and AGN bluer).  Chance coincidence probabilities can be tested by offsetting X-ray positions and checking the number of matches, allowing an estimate of the number of false matches \citep{Huang10,Lu09,Lan10}.   For individual sources, strong evidence for an AB nature could include the detection of H-$\alpha$ emission \citep{Cohn10}, location in the binary region (up to 0.75 magnitudes above the main sequence) \citep{Bassa08,Cohn10}, location in the ``red straggler'' region \citep{Heinke05a,Bassa08,Lu09,Huang10}, or a lack of any ultraviolet excess, as observed in all CVs from their hotter component(s) \citep{Edmonds03a}.  Large numbers of ABs have been clearly identified by these methods (see Fig. 4, left) where deep \Chandra\ and HST imaging is available for nearby clusters. 

A low X-ray/optical flux ratio ($<\sim1$)  is often used to rule out an AB nature for relatively X-ray bright stars \citep[e.g.][]{Kong06,Lu09,Lan10}. We test this with Fig. 4 (right), where we plot $L_X$ vs. $M_V$ for a large number of clearly identified X-ray counterparts in four well-studied globular clusters, omitting sources for which the X-ray to optical flux ratio was used in the classification (e.g. CX1 in M4, U18 in NGC 6397).  The diagonal lines indicate suggested dividing lines between CVs and ABs from \citet{Bassa04} and \citet{Verbunt08}.  Most identified ABs indeed lie below both lines, and most CVs above.  However, a couple clearly identified ABs appear above even \citet{Bassa04}'s suggested line; W47 in 47 Tuc \citep[a semidetached, near-contact binary according to its lightcurve;][]{Albrow01} is particularly X-ray luminous.  \citet{Edmonds03b} suggested that the higher X-ray to optical flux ratio of a few detected ABs in 47 Tuc, compared to nearby ABs studied by \citet{Dempsey97}, could simply be due to X-ray selection from a large binary population.  

\begin{figure}
  \includegraphics[height=.4\textheight]{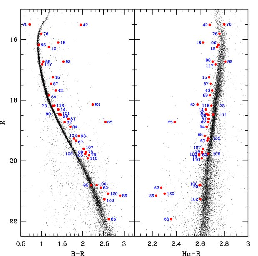}
  \includegraphics[height=.4\textheight]{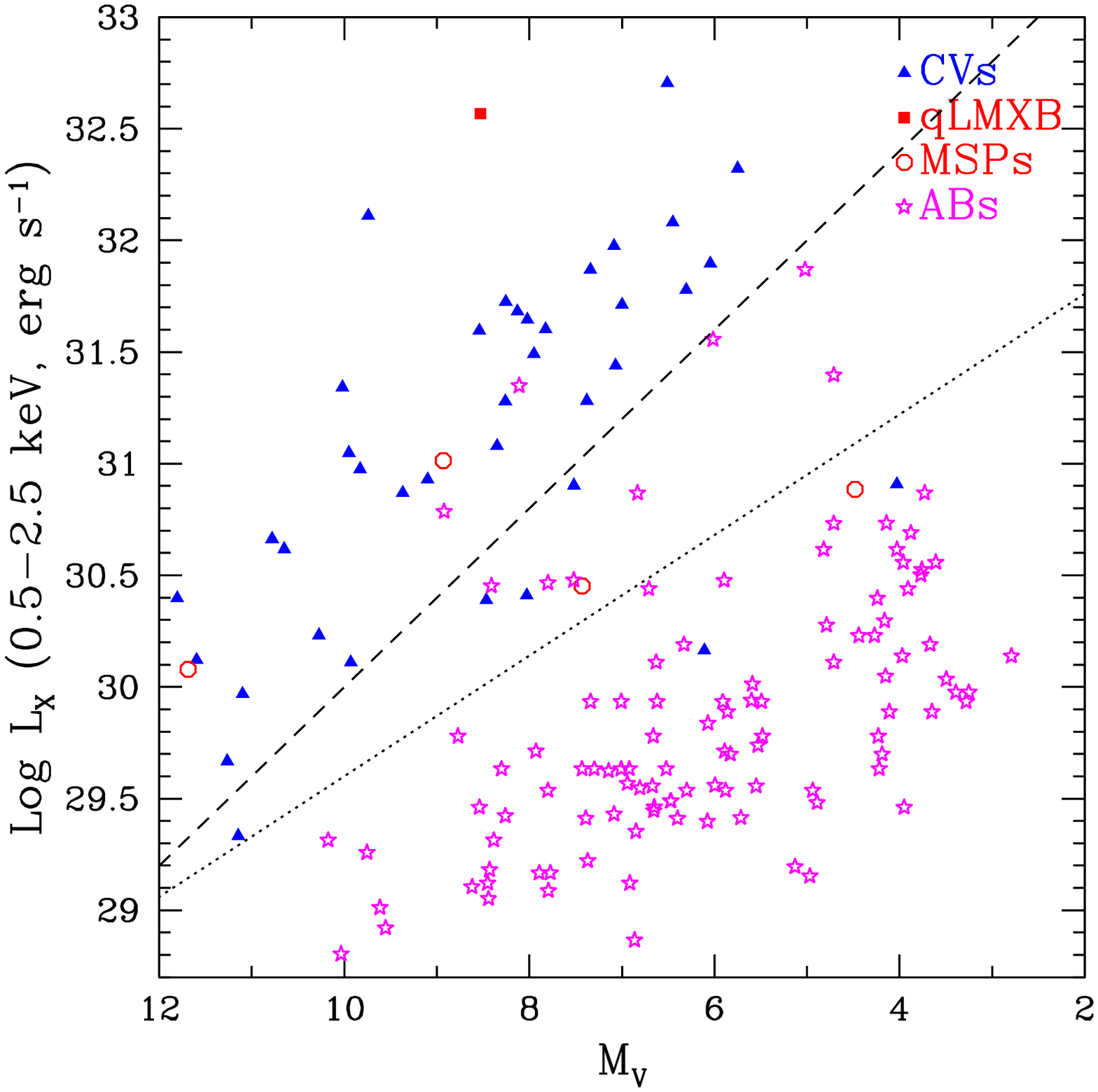}
  \caption{Left: CMDs of NGC 6397, marking identifications of ABs (along with the MSP U12 and candidate MSP U18), from \citet{Cohn10}.  The objects U42, U92, U63, U65, and U86 are probably foreground ABs.  Right: $L_X$ (0.5-2.5 keV) vs. $M_V$ for clearly identified X-ray sources in 47 Tuc \citep{Edmonds03a,Heinke05a, Albrow01}, NGC 6752 \citep{Pooley02a}, NGC 6397 \citep{Bogdanov10a,Cohn10}, and M4 \citep{Bassa04}.  Where $V$ observations aren't available, we assume $V$=($B$+$R$)/2.  Diagonal lines indicate suggested dividers between CVs and ABs from \citet[][upper, dashed]{Bassa04} and \citet[][lower, dotted]{Verbunt08}. 
}
\end{figure}

Unlike neutron star binaries, ABs should be common in primordial globular cluster populations.  \citet{Bassa04} presented evidence that the number of ABs scale with mass, rather than collision rate, among the globular clusters 47 Tuc, M4, and NGC 6397.  Such a scaling with mass suggests that their origin is indeed primordial.  This trend has been confirmed by identifications of ABs in relatively low-density clusters \citep{Kong06,Bassa08,Lu09,Lan10,Huang10}.  However, \citet{Verbunt99} showed that most globular clusters have lower overall X-ray to optical flux ratios than the open cluster M67, where ABs dominate the X-ray luminosity \citep[e.g.][]{vandenBerg04}.  We are now \citep{Heinke10c} confirming and extending that conclusion by showing that the X-ray to optical flux ratio of all old stellar populations (elliptical galaxies, dwarf spheroids, open clusters, the old stellar population in the Galactic disk) is much higher than the average flux ratio of low- to moderate-density globular clusters. In particular, the X-ray to optical flux ratio of the local old stellar population (and at least two open clusters) is dominated by ABs, indicating that the primordial AB population in all globular clusters is reduced.  This suggests that the AB population scales with the overall binary population, which is known to be much lower in globular clusters than other populations \citep[e.g.][]{Sollima10}.
Either globular clusters are born with fewer binaries, or many of their binaries are destroyed over time; further study of the dynamical evolution of globular clusters may shed light on this question \citep{Fregeau09}.

X-ray observations (particularly with \Chandra) of galactic globular clusters, combined with multiwavelength observations (particularly with \HST), have led to extraordinary progress in the identification and interpretation of X-ray sources in the last decade.  Broad studies of numerous clusters are defining the role of dynamics in producing cluster X-ray sources, while deep studies of selected clusters are producing a better understanding of entire X-ray source populations.  The scientific returns from globular cluster X-ray studies are continuing to grow.

%%%%%%%%%%%%%%%%%%%%%%%%%%%%%%%%%%%%%%%%%%%%%%%%
%% BACKMATTER
%%%%%%%%%%%%%%%%%%%%%%%%%%%%%%%%%%%%%%%%%%%%%%%%

\begin{theacknowledgments}
  I thank the collaborators whom I have worked on globular clusters with, especially J. Grindlay, P. Edmonds, H. Cohn, P. Lugger, R. Wijnands, D. Altamirano, S. Bogdanov, E. Cackett, F. Camilo, A. Cool, D. Lloyd, R. Elsner, P. Freire, R. Huang, N. Ivanova, A. Kong, W. Lewin, R. Narayan, D. Pooley, S. Ransom, G. Rybicki, and M. van den Berg.  I thank H. Cohn, P. Lugger, and S. Bogdanov for permission to use their figures. 
 I am grateful to NSERC and the Ingenuity New Faculty Award Program for financial support, and to the teams that maintain the observatories (especially \Chandra) that make this work possible.
\end{theacknowledgments}



\bibliographystyle{aipproc}   % if natbib is available

\bibliography{src_ref_list}

\IfFileExists{\jobname.bbl}{}
 {\typeout{}
  \typeout{******************************************}
  \typeout{** Please run "bibtex \jobname" to optain}
  \typeout{** the bibliography and then re-run LaTeX}
  \typeout{** twice to fix the references!}
  \typeout{******************************************}
  \typeout{}
 }

\end{document}